\documentclass[]{aa} 
\usepackage{graphicx,rotating}
\usepackage{txfonts}
\usepackage{soul}
\usepackage{natbib}
\usepackage{gensymb}
\usepackage{subfig}
\usepackage{color}
\begin{document}
\title{N$_2$H$^+$ and N$^{15}$NH$^+$ towards the prestellar core 16293E in L1689N}
\titlerunning{N$_2$H$^+$ in the prestellar core 16293E}
\authorrunning{Daniel et al.}
\author{
F.~Daniel\inst{\ref{inst0},\ref{inst1}} 
\and A. Faure\inst{\ref{inst0},\ref{inst1}} 
\and L. Pagani\inst{\ref{inst4}}
\and F. Lique\inst{\ref{inst7}} 
\and M. G\'erin\inst{\ref{inst3}}
\and D. Lis\inst{\ref{inst4},\ref{inst9}}
\and P. Hily--Blant\inst{\ref{inst0},\ref{inst1}} 
\and A. Bacmann\inst{\ref{inst0},\ref{inst1}} 
\and E. Roueff\inst{\ref{inst8}}
} 
\institute{
Univ. Grenoble Alpes, IPAG, F-38000 Grenoble, France  \label{inst0} 
\and
CNRS, IPAG, F-38000 Grenoble, France \label{inst1}  
\and
LERMA, Observatoire de Paris, PSL Research University, CNRS, Sorbonne Universit\'es, UPMC Univ. Paris 06, F-75014, Paris, France  \label{inst4}
\and
LOMC-UMR 6294, CNRS-Universit\'e du Havre, 25 rue Philippe Lebon, BP 1123 -- 76063 Le Havre Cedex, France \label{inst7} 
\and
LERMA, Observatoire de Paris, PSL Research University, CNRS, Sorbonne Universit\'es, UPMC Univ. Paris 06, Ecole normale sup\'erieure, F-75005, Paris, France  \label{inst3} 
\and
301-17, California Institute of Technology, Pasadena, CA 91125, USA \label{inst9}
\and
LERMA, Observatoire de Paris, PSL Research University, CNRS, Sorbonne Universit\'es, UPMC Univ. Paris 06, F-92190, Meudon, France  \label{inst8}  
}

\date{Received; accepted}


\abstract
{Understanding the processes that could lead to enrichment of molecules in
  $^{15}$N atoms is of particular interest in order to shed light on
  the relatively large variations observed in the $^{14}$N/$^{15}$N
  ratio in various solar system environments.}
{Currently, the sample of molecular clouds where $^{14}$N/$^{15}$N
  ratios have been measured is small and has
  to be enlarged in order to allow statistically significant studies.
  In particular, the N$_2$H$^+$ molecule currently shows the
  largest spread of $^{14}$N/$^{15}$N ratios in high-mass star forming
  regions. However, the $^{14}$N/$^{15}$N ratio in N$_2$H$^+$ was
  obtained in only two low-mass star forming regions (L1544 and B1b).
  The current work extends this sample to a third dark cloud.}
{We targeted the 16293E prestellar core, where the N$^{15}$NH$^+$
  $J$=1-0 line was detected. Using a model previously developed for
  the physical structure of the source, we solved the molecular
  excitation with a non-local radiative transfer code. For that purpose, we 
  computed specific collisional rate coefficients for the N$^{15}$NH$^+$-H$_2$
  collisional system. As a first step of the analysis, the N$_2$H$^+$ abundance profile was
  constrained  by reproducing the N$_2$H$^+$ $J$=1-0 and 3-2 maps. A scaling factor
  was then applied to this profile to match the
  N$^{15}$NH$^+$ $J$=1-0 spectrum.}
{We derive a column density ratio N$_2$H$^+$ / N$^{15}$NH$^+$ = $330^{+170}_{-100}$. }
{ We performed a detailed analysis of the excitation of N$_2$H$^+$ and
  N$^{15}$NH$^+$ towards the 16293E core, using state-of-the-art
  models to solve the radiative transfer, as well as the most accurate
  collisional rate coefficients available to date, for both
  isotopologues. We obtained the third estimate of the N$_2$H$^+$ /
  N$^{15}$NH$^+$ column density ratio towards a cold prestellar
  core. The current estimate $\sim$330 agrees with the value typical
  of the elemental isotopic ratio in the local ISM. It is however
  lower than in some other cores, where values as high as 1300 have been
  reported.  }


\maketitle
%

\section{Introduction}

Molecular isotopic ratios are invaluable tools for studying the origin of
solar system materials and their possible link with interstellar
chemistry. After hydrogen, the largest isotopic variations in the solar
system are observed for nitrogen. The $^{14}$N/$^{15}$N ratio varies
by a factor of $\sim$9, from the protosolar nebula (PSN) value of $\sim$440
to the ``hotspots'' in meteorites where ratios as low as $\sim$50
have been reported \citep[see][and references
  therein]{furi2015}. Enrichment in $^{15}$N with respect to the
PSN value is observed in most objects of
the solar system, except Jupiter. The reference value for
cosmochemists is that of Earth with a $^{14}$N/$^{15}$N ratio of 272
(measured in atmospheric N$_2$). In comets, the $^{14}$N/$^{15}$N
ratio has been measured in the three species : CN, HCN and NH$_2$, and
the measurements were shown to cluster near $^{14}$N/$^{15}$N$\sim$150
\citep[see][and references therein]{rousselot2014}.

Several hypotheses have been proposed to explain $^{14}$N/$^{15}$N enrichments in
the solar system, which fall into two main categories. The
first category concerns specific isotopic effects associated with the N$_2$
photodissociation by UV light from the protosun or from nearby stars,
such as self-shielding \citep[e.g.]{lyons2009}. In the second category, $^{15}$N
enrichment is caused by chemical fractionation through ion-molecule
reactions in the cold and dense interstellar medium (ISM), or in the
cold regions of the protosolar disk.

In the dense ISM, direct observation of the
nitrogen reservoir (presumably N or N$_2$) is not possible and the
$^{14}$N/$^{15}$N ratio has been obtained so far from the trace
species N$_2$H$^+$, NH$_3$, NH$_2$D, CN, HCN, and HNC. In the local
ISM, $^{15}$N enrichments ($^{14}$N/$^{15}$N $< 300$) have been
measured in CN, HCN and HNC
\citep{ikeda2002,adande2012,hilyblant2013a,wampfler2014}, but not in
the ammonia isotopologues for which the $^{14}$N/$^{15}$N ratio is
close to the PSN value or larger
\citep{gerin2009,lis2010,daniel2013}. The case of N$_2$H$^+$ is the
most intriguing, with values ranging from $\sim$180 to $\sim$1300
\citep{bizzocchi2013,daniel2013,fontani2015}. The observational
situation has thus significantly improved in the last five years, since
a number of new $^{14}$N/$^{15}$N estimates have been made available. 
The results, however, are still puzzling.
In fact, the large spread in molecular $^{14}$N/$^{15}$N ratios reflects, at least partly, the difficulty of
the measurements due to opacity and excitation effects, or the resort
to the double isotope method. In addition, a gradient of
$^{14}$N/$^{15}$N with the galactocentric distance has been measured
in CN and HNC by \citet{adande2012}, as predicted by galactic chemical
evolution models. The unambiguous observation of $^{15}$N chemical
fractionation in the ISM thus remains a challenging task. We also note
that recently, the $^{14}$N/$^{15}$N ratio has been determined for the
first time in a protoplanetary disk \citep{guzman2016}. A ratio of
200$\pm$100 was inferred from HCN, which is compatible with the
measurements previously reported in dark clouds and comets.

Turning to theory, the pioneering model of \citet{terzieva2000} predicted
that chemical fractionation in $^{15}$N should only be modest ($\sim$25\%)
 and hardly detectable in the ISM. Subsequently, a fractionation
mechanism based on CO depletion was suggested, predicting
that two different pathways can drive the $^{15}$N-fractionation: a
slow one to ammonia and a rapid one to HCN and other nitriles
\citep{charnley2002,rodgers2004,rodgers2008}. A chemical origin of
the differential $^{15}$N enrichment between hydrides and nitriles was
also proposed by \citet{hilyblant2013a,hilyblant2013b}.
In addition, by considering nuclear-spin effects in
ion molecule reactions involving the ortho and para forms of H$_2$,
\cite{wirstrom2012} have shown that the $^{15}$N enrichments of
nitriles do not correlate with deuterium (D) enrichments, as observed
in meteorites. On the other hand, a recent reinvestigation of
gas-phase chemical processes including D, $^{13}$C and $^{15}$N
species has suggested that the main $^{15}$N-fractionation routes are
in fact inefficient \citep{roueff2015}. Hence, though theory has 
improved, no model is currently able to reproduce the whole set of
observational data. It is generally believed that important routes
of nitrogen fractionation are still missing in the models
\citep[see, e.g.,][]{fontani2015}.

The dyazenilium ion (N$_2$H$^+$) is an interesting target for several
reasons. First, chemically, it is a direct daughter product of N$_2$,
one of the two main nitrogen reservoirs with atomic nitrogen, through
the proton transfer reaction N$_2$+H$_3^+\rightarrow$
N$_2$H$^+$+H$_2$. Second, the $^{14}$N/$^{15}$N ratio in N$_2$H$^+$
can be determined without recourse to the double isotope method, a method
commonly used for carbon bearing species 
\citep[see, e.g.,][]{adande2012,hilyblant2013a,hilyblant2013b}. Third,
accurate collisional rate coefficients for N$_2$H$^+$+H$_2$ have recently been
made available \citep[see][and below]{lique2015}. Finally,
observations of N$_2$H$^+$ show the largest variations in the
$^{14}$N/$^{15}$N ratio among all the nitrogen carriers. In the
prestellar core L1544, this ratio was estimated as $\sim$1000 $\pm$
200 \citep{bizzocchi2013} with N$^{15}$NH$^+$ and $^{15}$NNH$^+$
having nearly equal abundances. In the B1b molecular cloud, the
$^{14}$N/$^{15}$N ratio was estimated as $400^{+100}_{-65}$ in N$^{15}$NH$^+$
\citep{daniel2013}. The low signal-to-noise ratio achieved for the observations of 
$^{15}$NNH$^+$ isotopologue led to a lower limit of 600 for this
isotopologue. Finally, in high-mass star forming cores, the
$^{14}$N/$^{15}$N ratio was estimated in the range 180--1300
\citep{fontani2015}.

In the present work, we provide a new measurement of the
$^{14}$N/$^{15}$N ratio derived from observations of two
isotopologues of dyazenilium, N$_2$H$^+$ and N$^{15}$NH$^+$, towards
the low-mass prestellar core 16293E in L1689N. Note that contrary 
to the L1544 prestellar core, this source is not isolated and is influenced by 
the close--by Class 0 protostar IRAS 16293-2422.
Towards this region, previous observations of some of the N$_2$H$^+$ isotopologues 
were reported by \citet{castets2001} for N$_2$H$^+$ and \citet{gerin2001} for N$_2$D$^+$.
In Section~2, we give
details on the observations used in this study. Section~3 describes
the calculations performed for the N$^{15}$NH$^+$-H$_2$ rate
coefficients. Section~4 deals with the molecular excitation
calculations and our conclusions are given in Section~5.

\section{Observations}\label{sec:obs}

In the current study, we used the N$_2$H$^+$ $J$=1-0 SEST 15m telescope
observations reported in \citet{castets2001}. We discovered, however, that these
data suffered from observational artifacts, by comparing the
observed spectra with spectra of the same line observed more recently
at the IRAM 30m Telescope. This problem is further detailed below.

IRAM 30m observations of N$_2$H$^+$ were performed in early 2015
(17$^{th}$ of February and 2$^{nd}$ of April).  Weather was bad,
cloudy and unstable in February but excellent (1--2 mm precipitable
water vapor) in April. Saturn, only a few degrees away from the source,
was used for pointing and focusing of the telescope
during both runs.  Pointing was good with an uncertainty below 3$\arcsec$. Observations
were done with the EMIR0 receivers in frequency switch mode with both the
autocorrelator (VESPA) at 9.8 kHz (= 31.4 m\,s$^{-1}$) sampling and
the fast fourier transform spectrometer (FFTS) at 48.8 kHz (= 157
m\,s$^{-1}$) sampling.  The rest frequency of the line for the main
hyperfine component is 93173.764 MHz, following
\citet{pagani2009}. T$_\mathrm{sys}$ was typically 100 K (T$_a^*$
scale) with the source elevation ranging from 18.5$\degree$ to
28.5$\degree$. Each integration lasted one minute and both polarisations
were averaged and then folded. The typical noise is 70 mK
(after subtracting a second or third order baseline). The FWHM beam size is
26.5$\arcsec$ and the sampling was every 15$\arcsec$. The main beam
efficiency (0.80) is interpolated from an IRAM table (0.81 at 86 GHz and
0.78 at 115 GHz). These data will be presented and analysed in detail
in a forthcoming paper (Pagani et al., in prep.) along with other
observations.

Convolving the IRAM data to the SEST resolution, we find a relatively
good correspondence in peak intensity, but not in the line
profile. Consequently, the integrated intensity of the SEST data (16.4
$\pm$ 0.04 K km s$^{-1}$) is somewhat higher than the one from IRAM
(14.1$\pm$ 0.003 K km s$^{-1}$), i.e., +16\%. The discrepancy is much
larger than the systematic uncertainties.  The SEST data \citep{castets2001}
 were obtained with an acousto-optical spectrometer which was
the cause of an undue widening of the lines and line shape change,
which cannot be reproduced by simply smoothing the IRAM data. To use
the SEST data, we therefore multiplied them by 0.86 and
introduced an ad--hoc convolution in frequency, as described in the
next section.

Single-dish observations of the 1 mm molecular transitions presented
here were carried out in 2013 May--June, using the 10.4 m Leighton
Telescope of the Caltech Submillimeter Observatory (CSO) on Mauna Kea,
Hawaii. We used the wideband 230~GHz facility SIS receiver and the
FFTS backend that covers the full 4 GHz intermediate frequency (IF)
range with a 270~kHz channel spacing (0.37 km s$^{-1}$ at 220~GHz). Pointing
of the telescope was checked by performing five-point continuum scans
of planets and strong dust continuum sources. The CSO main-beam
efficiency at 230~GHz at the time of the observations was determined
from total-power observations of planets to be $\sim$65\%. The
absolute calibration uncertainty is $\sim$15\% and the FWHM CSO beam 
size is $\sim$35$\arcsec$ at 220~GHz.

The N$^{15}$NH$^+$ $J$=1-0 observations were performed in August 2014 with the IRAM-30m
Telescope during average weather conditions (2--4 mm of precipitable water
vapor).
The Eight Mixer Receiver (EMIR) receiver was tuned to 92.0 GHz.
We used the FFTS backend with 49.8 kHz ($\sim$160 m s$^{-1}$) spectral resolution, leading to an
instantaneous bandpass of 1.8 GHz and a velocity resolution of 0.16
kms$^{-1}$.
The $J$=1-0 transition of $^{15}$NNH$^+$, around 90.25 GHz, was not covered with the chosen
setup, which aimed at detecting other species. The system temperature
was about 140 K and the source elevation varied between 18$^\circ$ and 28$^\circ$.
The observations were performed in position switching, with the
reference position set 300$\arcsec$ west of the source. The nearby planets
Mars and Saturn were used for checking the telescope pointing and the focus.
The data were reduced with CLASS and the noise level is 19 mK after removal of 
linear base lines.

\section{N$_{2}$H$^{+}$ and N$^{15}$NH$^{+}$ collisional rate coefficients} \label{sec:rates}

Collisional rate coefficients for the
N$_{2}$H$^{+}$-H$_2$($J$=0)\footnote{In molecular dynamics, the
  rotational quantum numbers are noted $j$, and $J$ refers to the
  total angular momentum. However, in what follows, we use the
  spectroscopic notation and denote the rotational quantum number as
  $J$. This choice is made for consistency with the other parts of the
  article since the spectroscopist notation is usually adopted in
  astrophysical studies.} system have been published recently by
\cite{lique2015}.  Hyperfine-structure-resolved excitation rate
coefficients, based on a new potential energy surface (PES) obtained
from highly correlated {\it ab initio} calculations
\citep{Spielfiedel:15}, were calculated for temperatures ranging from
5 K to 70 K. The new rate coefficients are significantly larger than the
N$_2$H$^+$-He rate coefficients previously published
\citep{Daniel:05}. In addition, the differences cannot be reproduced
by a simple scaling relationship.

As a first approximation, these new rate coefficients could be used to
analyse N$^{15}$NH$^{+}$ emission spectra since both N$_2$H$^+$ and
N$^{15}$NH$^+$ share the same molecular properties (when the hyperfine
splitting induced by the internal nitrogen is neglected). 
Recent studies, however, have shown that isotopic effects in inelastic
collisions can be important \citep{scribano2010,Dumouchel:12}, even in
the case of $^{14}$N$\to$$^{15}$N substitution
\citep{Flower:15}. Hence, we have decided to compute specific
N$^{15}$NH$^{+}$-H$_2$($J$=0) rate coefficients.

Within the Born-Oppenheimer approximation, the full electronic ground
state potential is identical for the N$_{2}$H$^{+}$-H$_2$ and
N$^{15}$NH$^{+}$-H$_2$ systems and depends only on the mutual
distances of the five atoms involved. Then, we used for the scattering
calculations the N$_{2}$H$^{+}$-H$_2$ PES of \cite{Spielfiedel:15} and
the ``adiabatic-hindered-rotor'' treatment, which allows
para-H$_2$($J$=0) to be treated as if it was spherical. Zero-point
vibrational effects are different in N$_{2}$H$^{+}$ and
N$^{15}$NH$^{+}$, but these effects are expected to be moderate and are
hence neglected in the present calculations.  The only difference
between the N$_{2}$H$^{+}$-H$_2$ and N$^{15}$NH$^{+}$-H$_2$ PES is
thus the position of the center of mass taken for the origin of the
Jacobi coordinates, an effect which has been taken into account in our calculations.
 
Since the $^{14}$N and $^{15}$N nitrogen atoms possess a non-zero
nuclear spin ($I$=1 and $I$=1/2, respectively), the N$_2$H$^+$ and
N$^{15}$NH$^{+}$ rotational energy levels are split into hyperfine
levels. In the astronomical observations, however, the hyperfine
structure due to $^{15}$N is not resolved. The hyperfine levels of the
N$^{15}$NH$^{+}$ molecules are thus characterized by the two quantum
numbers $J$ and $F$. Here, $F$ results from the coupling of $\vec{J}$
with $\vec{I}$ ($I$ being the nuclear spin of the external $^{14}$N
atom).

The hyperfine splitting in the N$_2$H$^+$ isotopologues is very
small. Assuming that the hyperfine levels are degenerate, it is
possible to simplify the hyperfine scattering problem using recoupling
techniques as described in \cite{Daniel:04,Daniel:05} and in \citet{Faure:12}. Hence,
we performed Close-Coupling calculations \citep{Arthurs:60} for the
pure rotational excitation cross--sections using the MOLSCAT program
\citep{molscat:94}, as in \cite{lique2015} for the main
isotopologue. The N$^{15}$NH$^{+}$ rotational energy levels were
computed using the rotational constants of
\cite{Dore:09}. Calculations were carried out for total energies up to
500~cm$^{-1}$. Parameters of the integrator were tested and adjusted
to ensure a typical precision to within 0.05 \AA$^2$ for the inelastic
cross sections. At each energy, channels with $J$ up to 28 were
included in the rotational basis to converge the calculations for all
the transitions between N$^{15}$NH$^{+}$ levels up to $J$=7. Using the
recoupling technique and the stored S-matrix elements, the
hyperfine-state-resolved cross--sections were obtained for all
hyperfine levels up to $J$=7.

From the calculated cross--sections, one can obtain the corresponding
thermal rate coefficients at temperature $T$ by an average over the
collision energy ($E_c$):

\begin{eqnarray}
\label{thermal_average}
k_{\alpha \rightarrow \beta}(T) & = & \left(\frac{8}{\pi\mu k^3_{B} T^3}\right)^{\frac{1}{2}}  \times  \int_{0}^{\infty} \sigma_{\alpha \rightarrow \beta}(E_c) \, E_{c}\, e^{-\frac{E_c}{k_{B}T}}\, dE_{c}
\end{eqnarray}
where $\sigma_{\alpha \to \beta}$ is the cross--section from initial
level $\alpha$ to final level $\beta$, $\mu$ is the reduced mass of
the system, and $k_{B}$ is Boltzmann's constant. 
Using the computational scheme described above, we have obtained
N$^{15}$NH$^{+}$-H$_2$($J$=0) rate coefficients for temperatures up to
70~K. These coefficients should have an accuracy similar to those
computed for the main isotopologue by \cite{lique2015}. The main
source of uncertainty is thus the use of the adiabatic-hindered-rotor
approximation, which was shown to introduce errors below 5-10\% \citep{lique2015,Spielfiedel:15}. A
typical accuracy of 10\% is thus expected for the rate coefficients of
the N$_2$H$^+$ isotopologues. The complete set of (de-)excitation rate
coefficients with $J,\,J'\le 7$ will be made available through the
LAMDA \citep{schoier:05} and BASECOL \citep{Dubernet:13} databases.

\begin{figure}
\begin{center}
\includegraphics[width=8.0cm,angle=0.]{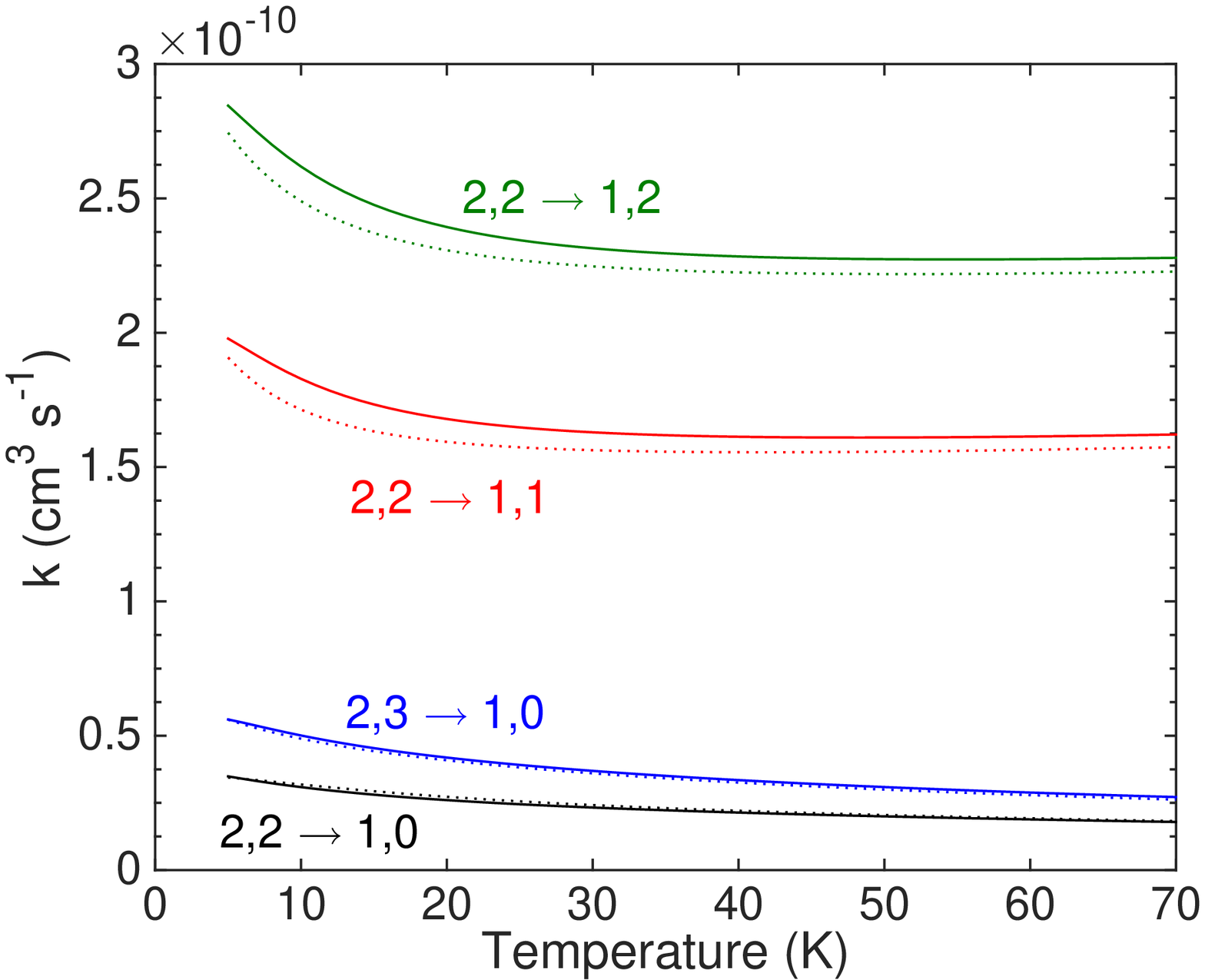}
\includegraphics[width=8.0cm,angle=0.]{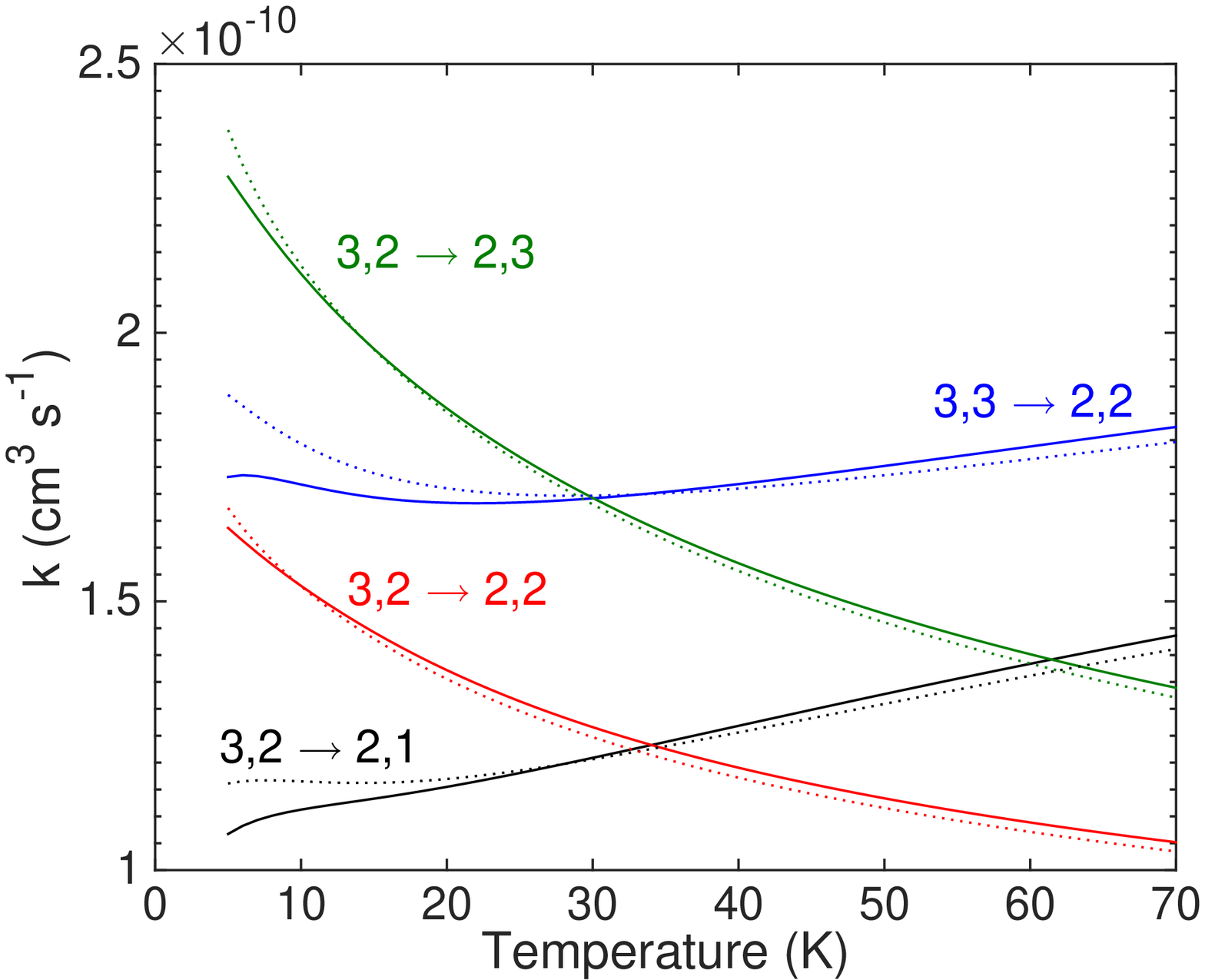}
\caption{Temperature variation of the hyperfine resolved
N$_2$H$^+$--H$_2(j=0)$ (solid lines) and N$^{15}$NH$^{+}$--H$_2(j=0)$ (dotted lines) rate coefficients for $j=2,F \to j'=1,F'$ and $j=3,F \to j'=2,F'$.
transitions.} \label{fig:rates}
\end{center}
\end{figure}

Figure \ref{fig:rates} presents the temperature variation of the
N$_2$H$^+$-H$_2$($J$=0) and N$^{15}$NH$^{+}$-H$_2$($J$=0) rate
coefficients for a few hyperfine transitions associated with the
$J$=3$\to$2 and 2$\to$1 rotational transitions. To enable a
direct comparison, we summed the N$_2$H$^+$-H$_2$($J$=0) rate
coefficients over the hyperfine structure associated with the internal
nitrogen nucleus.
As one can see, the differences between the N$_2$H$^+$ and
N$^{15}$NH$^{+}$ rate coefficients are moderate. The two sets of data
differ by less than 20 percent over the whole temperature
range. We find, however, that the largest deviations occur at
temperatures typical of cold molecular clouds ($T$=5--20 K) and we
observe that the N$_2$H$^+$ over N$^{15}$NH$^{+}$ rate coefficients
ratios tend to increase with decreasing temperature. The differences are
due to both the centre-of-mass shift in the interaction potential and
the use of a specific description of the isotopologue energy
levels. In conclusion, we found that the isotopologue specific
results differed from each other to an extent that they may be significant
for analyses of observations of the hyperfine transitions of these
species, at least for cold dark cloud conditions. 
Modeling N$^{15}$NH$^+$ with 
N$_2$H$^+$ rate coefficients will typically induce errors in the N$^{15}$NH$^+$ column density estimate
of the same order of magnitude as the differences found in rate coefficients. Hence, we can expect that earlier studies that resorted to 
the same set of rates for both isotopologues would not suffer from errors larger than 20\%. In particular, we expect this conclusion 
to remain true in the case where He is the collisional partner \citep{Daniel:05}. 

\section{Radiative transfer modeling} \label{sec:model}

To obtain an accurate estimate of the $^{14}$N/$^{15}$N ratio
in N$_2$H$^+$, we obtained the column densities of the two
isotopologues by solving the molecular excitation problem.  The
methodology used is similar to the approach described in
\citet{daniel2013}, where the $^{14}$N/$^{15}$N ratio was estimated
for various molecules towards B1b. In the present case, our analysis
deals with the 16293E prestellar core. To perform the
analysis, we have used the physical structure of 16293E described in
\citet{bacmann2015}. In that study, variations of the H$_2$
density and dust/gas temperature throughout the core were derived
using continuum observations at wavelengths ranging from 160 $\mu$m to
1.3 mm.  The core center was then fixed at an intermediate distance
between the maxima of the 850 $\mu$m and 1.3 mm maps. More precisely,
its coordinates are fixed at $\alpha$ = 16$^h$32$^m$28.8$^s$, $\delta$
= -24$\degr$29$\arcmin$4$\arcsec$ (J2000). In what follows, the offsets indicated in 
the figures are given according to this reference position.

The N$_2$H$^+$ spectroscopy is taken from \citet{caselli1995} and \citet{pagani2009}.
For the main N$_2$H$^+$ isotopologue, the N$_2$H$^+$-H$_2$ rate coefficients are taken from \citet{lique2015}.
For the rare N$^{15}$NH$^+$ isotopologue, the rate coefficients were described in the previous section and the spectroscopy
is taken from \citet{Dore:09}.
Finally, the molecular excitation and radiative transfer are solved with the \texttt{1Dart} code described in \citet{daniel2008},
which takes into account the line overlap between hyperfine lines. \\

\begin{figure}
\begin{center}
\includegraphics[angle=0,scale=.35]{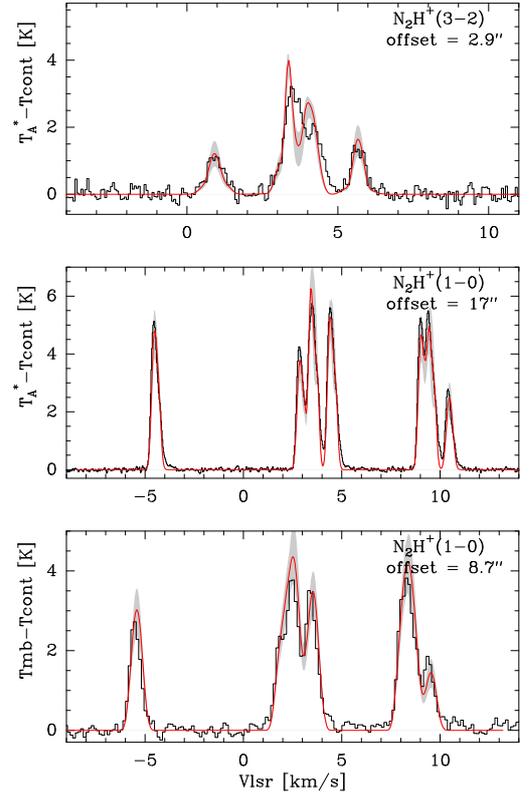}
\caption{Comparison between model and observations for the SEST (bottom panel) and IRAM (middle panel)
N$_2$H$^+$ $J$=1-0 line. The top panel shows the CSO N$_2$H$^+$ $J$=3-2 line.
In each panel, we give the offset between the position observed and the core center. For each spectrum, the grey area indicates the variation
of intensity obtained by varying the N$_2$H$^+$ abundance as is indicated in Fig. \ref{fig:profil_N2H+}} \vspace{-0.1cm}
\label{fig:central}
\end{center}
\end{figure}

As a first step of our analysis, we constrained the abundance of the
main N$_2$H$^+$ isotopologue throughout the core. To that purpose, we
made use of the $J$=1-0 and $J$=3-2 maps, respectively observed at the
SEST and CSO telescopes. Additionally, we used a spectrum of the $J$=1-0 line,
observed at the IRAM 30m Telescope towards a position offset by $\sim$17" from the core center. 
The comparison between the IRAM and SEST observations made by degrading the IRAM data
to the SEST resolution show
that the spectra obtained at the SEST telescope have broader
linewidths (see Section~2). To explain these differences, we assumed
that the SEST observations suffered from systematic uncertainties and
we thus only took into account the relative variations of the spectra
from one map position to another. Additionally, to reproduce
the observational artifact linked to the SEST observations, we
performed a spectral convolution of the synthetic spectra, with a
spectral response given by a Gaussian. The associated width was
adjusted so that a model that would fit the IRAM 30m $J$=1-0 spectra would
give an equally good fit for the closest SEST observation.
The corresponding point of the SEST map is 8.7$\arcsec$ from the core center
and the IRAM 30m and SEST $J$=1-0 observations are distant by $\sim$10$\arcsec$ from each other. 
The result of this procedure corresponds to the observations plotted in the bottom and middle
panels of Fig.~\ref{fig:central}.

\begin{figure*}
\begin{center}
\includegraphics[angle=0,scale=.7]{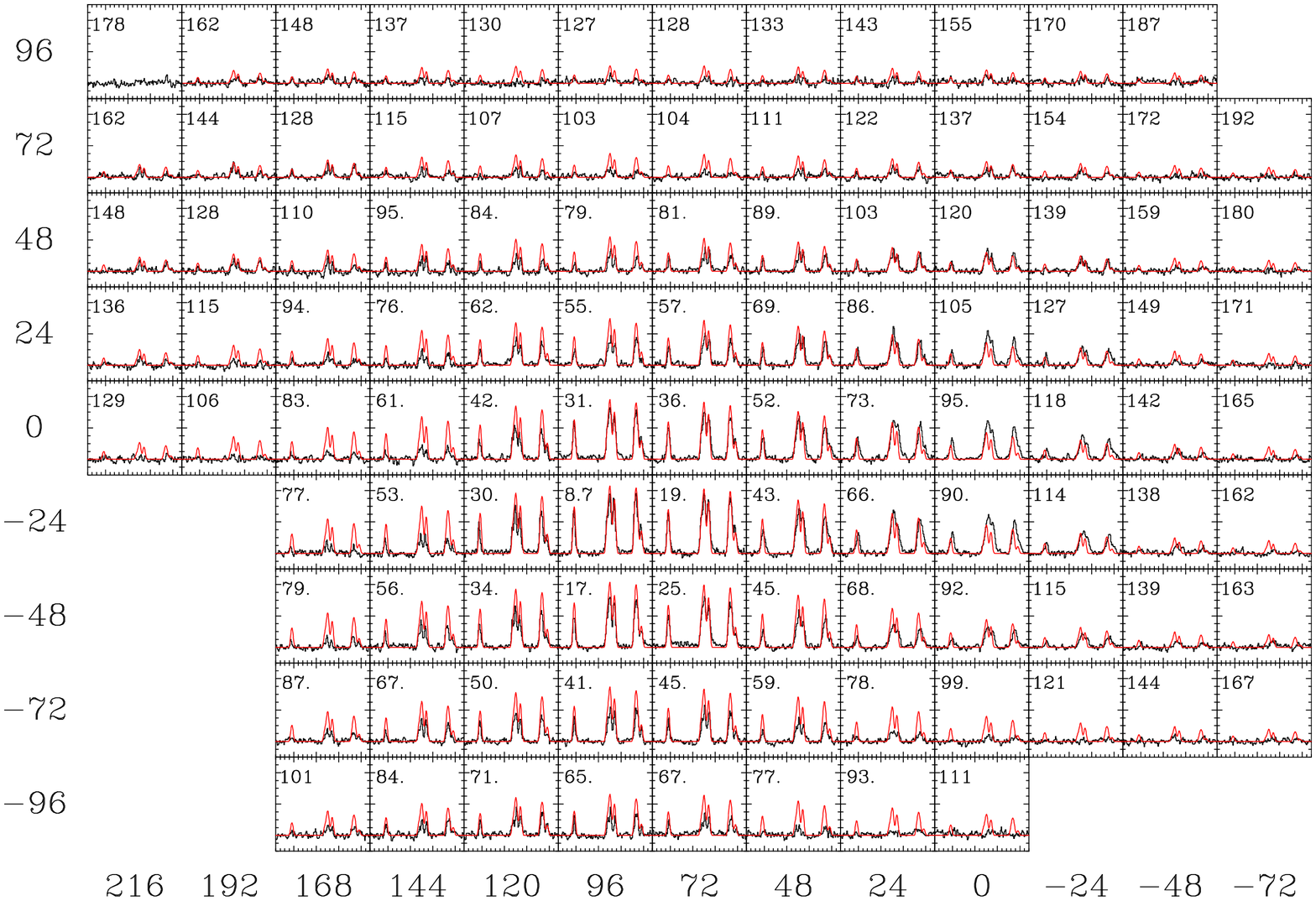}
\caption{Comparison between model and observations for the N$_2$H$^+$(J=1-0) map observed at SEST.
In each panel, the distance to the core center is indicated, in arcseconds, in the upper left side of the box. The reference
coordinates of the map is $\alpha$ = 16$^h$32$^m$22.76$^s$,  $\delta$ = -24$\degr$28$\arcmin$33.1$\arcsec$ (J2000), 
as given in \citet{castets2001}, which corresponds to a position close to the 16293-2422 protostar. This protostar is
at $\sim$90" at the North--West of the 16293E core center.} \vspace{-0.1cm}
\label{fig:map_SEST}
\end{center}
\end{figure*}

\begin{figure*}
\begin{center}
\includegraphics[angle=0,scale=.7]{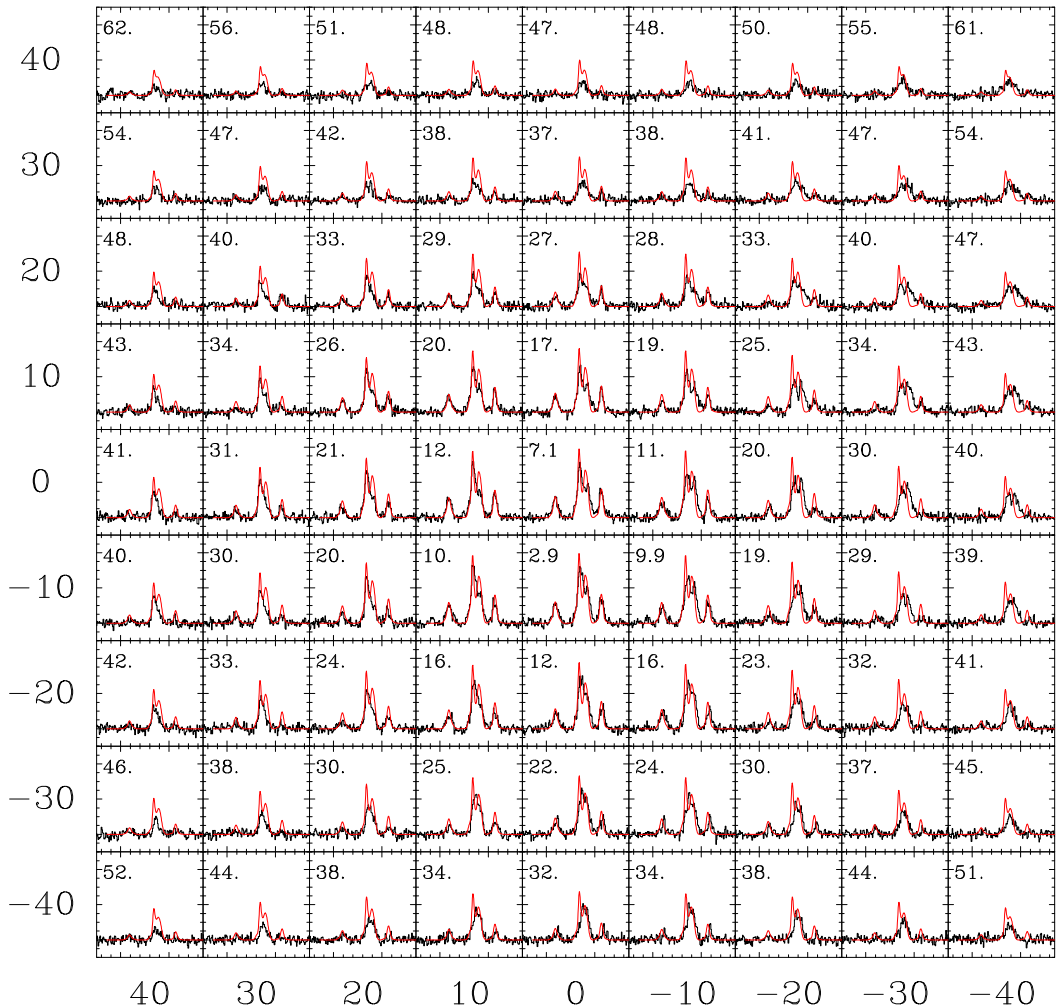}
\caption{Comparison between model and observations for the N$_2$H$^+$(J=3-2) map observed at CSO.
In each panel, the distance to the core center is indicated, in arcseconds, in the upper left side of the box.
The reference coordinates of the map is $\alpha$ = 16$^h$32$^m$28.83$^s$,  $\delta$ = -24$\degr$28$\arcmin$56.9$\arcsec$ (J2000),
which is at $\sim$7" from the position we assume for the core center.
} \vspace{-0.1cm}
\label{fig:map_CSO}
\end{center}
\end{figure*}

\begin{figure}
\begin{center}
\includegraphics[angle=0,scale=.35]{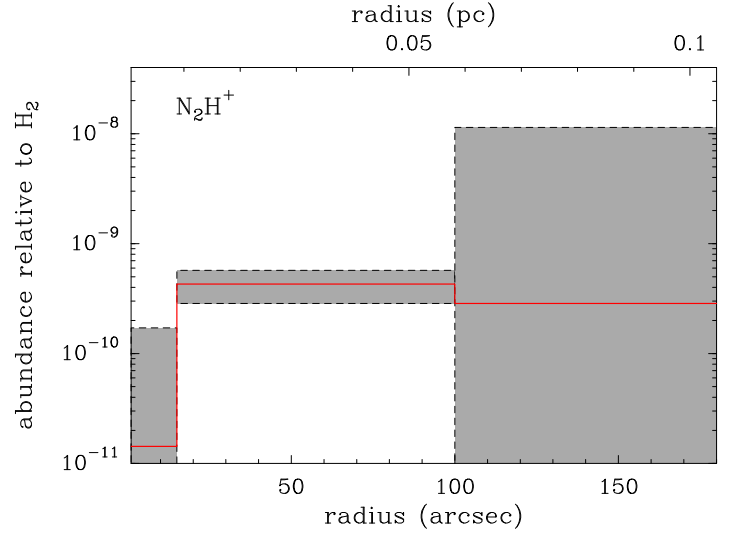}
\caption{N$_2$H$^+$ abundance profile.} \vspace{-0.1cm}
\label{fig:profil_N2H+}
\end{center}
\end{figure}

As previously said, the modeling is performed using the physical
structure (i.e., H$_2$ density, gas and dust temperatures, ...)
described in \citet{bacmann2015}. As a reminder, the density is constant at
$\sim$$1.4 \times 10^7$ cm$^{-3}$ within a radius of 4" and then decreases outward. The 
temperature at the core center is 11 K and increases outwards up to 16 K.
The only free parameters of the current modeling are thus linked to the N$_2$H$^+$
abundance profile. To fit the observations, we introduced
radial zones within which the N$_2$H$^+$ abundance is kept
constant. We started with the simplest model, i.e., with a constant
abundance throughout the cloud (i.e. one free parameter). We then
increased the number of radial zones, i.e., the number of free
parameters, until we obtained a reasonable fit to the
observations, the quality of the fit being gauged by eye. Doing so, a satisfactory fit is obtained with only three
radial regions. The fit of the $J$=1-0 and 3-2 lines obtained at
positions close to the core center are shown in
Fig. \ref{fig:central}, and the SEST and CSO maps are shown in
Fig. \ref{fig:map_SEST} and Fig. \ref{fig:map_CSO}. As can be seen in
these Figures, some discrepancies exist between the model and
observations at some particular positions. 
We estimate, however, that the overall agreement is satisfactory given the 
non--sphericity of the 16293E core \citep[see the molecular emission maps in, e.g.,][]{castets2001,lis2002b}.  The
N$_2$H$^+$ abundance that corresponds to these observations is
reported in Fig. \ref{fig:profil_N2H+}. In each region of the abundance profile, we determined error bars by varying the 
N$_2$H$^+$ abundance so that the resulting spectra did not depart significantly from the best model. The corresponding
variations from our best estimate correspond to the grey zones in Fig. \ref{fig:central}.
The column density inferred from this model towards the center of the sphere is N(N$_2$H$^+$) =
$4.6_{-1.2}^{+6.0}\,10^{13}$ cm$^{-2}$.

\begin{figure}
\begin{center}
\includegraphics[angle=0,scale=.35]{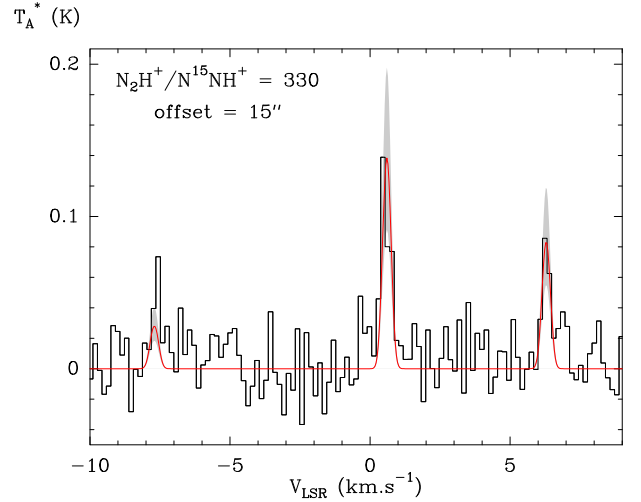}
\caption{Comparison between model and observations for the N$^{15}$NH$^+$ $J$=1--0 line.} \vspace{-0.1cm}
\label{fig:N15NH+_model}
\end{center}
\end{figure}

To model the spectrum of the N$^{15}$NH$^+$ $J$=1-0 line 
observed at $\sim$15'' from the core center, we assumed that the abundance profile is
similar to that of the main isotopologue. The observed spectrum is
then reproduced by introducing an overall scaling factor to the
abundance.  The observations are correctly reproduced with an
abundance ratio N$_2$H$^+$ / N$^{15}$NH$^+$ $\sim$$330^{+170}_{-100}$
and the comparison between the model and observations is shown in
Fig. \ref{fig:N15NH+_model}. In this Figure, the grey area corresponds to the 
error bars of the ratio. 

As in \citet{daniel2013}, we can calculate a mean excitation temperature $\bar{T}_{ex}$ from the source model.
For the N$_2$H$^+$ and N$^{15}$NH$^+$ J=1-0 lines, we respectively derive values of $9.44^{+0.97}_{-1.13}$ K and
$9.48^{+0.01}_{-0.01}$ K, where the sub-- and superscripts indicate the spread of values over all the hyperfine components.
The larger spread of values for the main isotopologue is a consequence of larger line opacities which is accompanied by
a departure from a single excitation temperature \citep{daniel2006}.
The opacities summed over all the components are respectively $\sim$17.3 and 0.06 for the N$_2$H$^+$ 
and N$^{15}$NH$^+$ isotopologues. Note that despite the higher opacity of the N$_2$H$^+$ $J$=1-0 line, 
$\bar{T}_{ex}$ is not  enhanced by line trapping effects for this isotopologue. In fact, in the current case, 
the increase in $T_{ex}$ due to line trapping is counter--balanced by the increase of the N$^{15}$NH$^+$-H$_2$ rate coefficients. 
Indeed, in the model of B1b, \citet{daniel2013} found that the $\bar{T}_{ex}$ of N$_2$H$^+$ was enhanced by $\sim$17\% with
respect to the N$^{15}$NH$^+$ $\bar{T}_{ex}$, the two calculations being performed with the same set of rate coefficients.
Hence, in the current case, the assumption of a similar $T_{ex}$ to describe the two isotopologues would apply.

\section{Discussion} \label{sec:discussion}
 
As explained in the Introduction, the large variations of the
$^{14}$N/$^{15}$N ratio in the solar system are not understood. As a
reminder, this ratio is $\sim$440 in the solar wind, $\sim$270 on
Earth, and $\sim$150 in comets. We refer the reader to \citet{furi2015}
for a recent review of the literature. The main question is whether
the $^{15}$N enrichments indicate interstellar inheritance, nitrogen
fractionation in the protoplanetary disk phase, or both. \citet{guzman2016} showed
recently that the average $^{14}$N/$^{15}$N ratio in HCN towards the
MWC 480 disk is 200$\pm 100$. This value is comparable to the ratios
observed in comets and in dark clouds, but the signal-to-noise ratios of those data
was not high enough to allow a pre-stellar or protoplanetary disk origin
to be distinguished \citep[see the discussion in][]{guzman2016}. In
addition, as explained above, $^{15}$N enrichments derived in HCN are
ambiguous due to the possible depletion of $^{13}$C
\citep{roueff2015}.

In the present work, we have provided a new measurement of the
$^{14}$N/$^{15}$N ratio in a prestellar core by observing N$_2$H$^+$
and N$^{15}$NH$^+$. The derived abundance ratio N$_2$H$^+$ /
N$^{15}$NH$^+$ $\sim$$330^{+170}_{-100}$ is comparable to the
elemental isotope ratio inferred for the local ISM, which is estimated
as $^{14}$N/$^{15}$N = 290$\pm$40 by \citet{adande2012}, from 
a survey of CN and HCN rotational lines. More recently, \citet{ritchey2015}
derived a local ISM value $^{14}$N/$^{15}$N = 274$\pm$18 from UV absorption lines of CN towards
four lines of sight.
Given the error bars, a ratio of $\sim$300 would apply to the 16293E region. 
This similarity to local ISM values would imply an absence
of chemical fractionation for N$_2$H$^+$ in this source, in dense gas. 
Indeed, the most recent gas-phase network of \cite{roueff2015} suggests that 
the fractionation reaction of $^{15}$N with N$_2$H$^+$ is
inefficient due to the presence of an activation barrier, in contrast to the
hypothesis made in previous models. In particular, the network of
\cite{hilyblant2013b} does predict $^{15}$N enrichment in N$_2$H$^+$, in
very good agreement with the present observation, but assumes a solar
elemental ratio of 440. An $^{14}$N/$^{15}$N value of $\sim$300, however, might be more appropriate to describe the 
local ISM. In that case, the difference with the PSN value then
would be a consequence of the stellar nucleosynthesis that enriched the ISM in $^{15}$N atoms by $\sim$50\%,
during the last 4.5 Gy \citep{ritchey2015}.

More observations are clearly needed to
establish the actual elemental ratio in 16293E and to confirm or
exclude nitrogen chemical fractionation in N$_2$H$^+$. In particular, observing various
$^{15}$N--substituted molecules is important since it would allow us to disentangle between
fractionation effects or a variation of the elemental abundance ratio from the mean local ISM value.
Indeed, in the first case, we expect to obtain a spread in the ratios, while in the second case, 
all the isotopic ratios should cluster around the same value.

More generally, it should be noted that gas-phase models of chemical
fractionation are dependent on the temperature, on the chosen
elemental abundances (especially the C/O ratio) and on the
ortho--to--para ratio of H$_2$ \citep{hilyblant2013b,legal2014,roueff2015}. It is also crucial to couple C, N and O
isotopic chemistries, as emphasized by \cite{roueff2015}. As a result, small
variations in chemical conditions can have a strong impact on the
predicted molecular $^{14}$N/$^{15}$N ratios. We note in this context
that the temperature at the core center is higher in 16293E
($\sim$11~K), where N$_2$H$^+$ / N$^{15}$NH$^+\sim$330, than in L1544
($\sim$6~K), where N$_2$H$^+$ / N$^{15}$NH$^+\sim$1000. It is unclear
whether such a small temperature difference can play a role but this
should be investigated in detail in future dedicated studies. The
large spread observed in high-mass star forming regions reported by
\citet{fontani2015} (N$_2$H$^+$ / N$^{15}$NH$^+\sim$180-1300) could
also reflect temperature or elemental abundance effects. It could
also be partly due to the gradient of the $^{14}$N/$^{15}$N abundance
ratio with galactocentric distance \citep{adande2012}. In any case,
the statistics of objects with a determination of the $^{14}$N/$^{15}$N ratio in various
molecules has to be enlarged if we want to answer the question of the
origin of nitrogen fractionation in the solar system.

\begin{acknowledgements}
This work has been supported by the Agence Nationale de la Recherche
(ANR-HYDRIDES), contract ANR-12-BS05-0011-01 and by the CNRS national
program ``Physico-Chimie du Milieu Interstellaire''. 
This work is based upon observations with the the Caltech Submillimeter Observatory, operated by the California Institute of Technology.
Support for this work was provided by NASA through an award issued by JPL/Caltech.
The authors thank A. Castets
for providing the N$_2$H$^+$ data acquired with the SEST telescope. 
\end{acknowledgements}

\bibliographystyle{aa}
\bibliography{biblitex}

\end{document}